\documentclass[11pt,letterpaper]{article}
\pdfoutput=1
\usepackage{jcappub}
\usepackage{color}
\usepackage{graphicx}
\usepackage{wrapfig}

\usepackage{cancel}

\usepackage{verbatim}
\usepackage{amsmath}
\usepackage{amssymb}
\usepackage{subfig}
\usepackage{url}
\usepackage{bbold}
\usepackage{xspace}

\usepackage{multirow}
\usepackage{threeparttable}
\usepackage{paralist}

\usepackage{color}
\definecolor{darkgreen}{rgb}{0,0.5,0}

\DeclareRobustCommand{\Sec}[1]{Sec.~\ref{#1}}

\DeclareRobustCommand{\App}[1]{App.~\ref{#1}}

\DeclareRobustCommand{\Fig}[1]{Fig.~\ref{#1}}

\DeclareRobustCommand{\Eq}[1]{Eq.~(\ref{#1})}
\DeclareRobustCommand{\Eqs}[2]{Eqs.~(\ref{#1}) and (\ref{#2})}
\DeclareRobustCommand{\Ref}[1]{Ref.~\cite{#1}}

\newcommand{\be}{\begin{equation}}
\newcommand{\ee}{\end{equation}}

\newcommand{\mb}[1]{\boldsymbol{#1}}

\def\ltap{\ \raise.3ex\hbox{$<$\kern-.75em\lower1ex\hbox{$\sim$}}\ }
\def\gtap{\ \raise.3ex\hbox{$>$\kern-.75em\lower1ex\hbox{$\sim$}}\ }
\def\lsim{\ \raise.3ex\hbox{$<$\kern-.75em\lower1ex\hbox{$\sim$}}\ }
\def\gsim{\ \raise.3ex\hbox{$>$\kern-.75em\lower1ex\hbox{$\sim$}}\ }
\def\eg{{\it e.g.}}
\def\ie{{\it i.e.}}

\usepackage{color}

\begin{document}
\begin{flushright}
FERMILAB-PUB-14-004-T\\
MIT-CTP-4528
\end{flushright}
\title{Taking Halo-Independent Dark Matter Methods Out of the Bin}

\author[a]{Patrick J. Fox,}

\affiliation[a]{Theoretical Physics Department, Fermilab, Batavia, Illinois 60510, USA}

\author[b]{Yonatan Kahn}

\author[b]{and Matthew McCullough}

\affiliation[b]{Center for Theoretical Physics, Massachusetts Institute of Technology,\\Cambridge, MA 02139, USA}

\emailAdd{pjfox@fnal.gov}
\emailAdd{ykahn@mit.edu}
\emailAdd{mccull@mit.edu}

\date{\today}

\keywords{}

\arxivnumber{}

\abstract{We develop a new halo-independent strategy for analyzing emerging DM hints, utilizing the method of extended maximum likelihood.  This approach does not require the binning of events, making it uniquely suited to the analysis of emerging DM direct detection hints.  It determines a preferred envelope, at a given confidence level, for the DM velocity integral which best fits the data using all available information and can be used even in the case of a single anomalous scattering event.  All of the halo-independent information from a direct detection result may then be presented in a single plot, allowing simple comparisons between multiple experiments.  This results in the halo-independent analogue of the usual mass and cross-section plots found in typical direct detection analyses, where limit curves may be compared with best-fit regions in halo-space.  The method is straightforward to implement, using already-established techniques, and its utility is demonstrated through the first unbinned halo-independent comparison of the three anomalous events observed in the CDMS-Si detector with recent limits from the LUX experiment.}

\maketitle

\section{Introduction}
\label{sec:introduction}

Despite the extraordinary progress in understanding the fundamental forces and building blocks of the Universe, the particle nature of dark matter (DM) remains unknown.  Unravelling this puzzle remains one of the key tasks facing particle physics.  Among the many facets of this wide-ranging endeavor, interactions between familiar matter and DM particles are searched for in underground direct detection experiments.  Specifically these experiments search for the energy which would be deposited when a DM particle strikes a nucleus.  Progress in this direction has been rapid and experiments have been extremely successful in pushing to ever lower interaction strengths, providing much needed knowledge about this aspect of DM phenomenology.

A few experiments have observed some potential signals of DM scattering, such as the long-standing DAMA annual modulation \cite{Bernabei:2010mq}, the CoGeNT excess and modulation \cite{Aalseth:2011wp,Aalseth:2014nda}, CRESST-II excess \cite{Angloher:2011uu}, and most recently the CDMS-Si excess \cite{Agnese:2013rvf}.  However, results from other experiments which have not observed any excess over expected backgrounds have increasingly put tension on DM interpretations of the positive hints, with the recent result from the LUX experiment excluding the simplest possibility, spin-independent elastically scattering DM where the DM couples equally to protons and neutrons \cite{Akerib:2013tjd,Gresham:2013mua,DelNobile:2013gba,Fox:2013pia}.\footnote{Further analysis of the LUX results reveal that DM interpretations of the CDMS-Si excess with unequal DM couplings to protons and neutrons \cite{Feng:2013vod,Gresham:2013mua,DelNobile:2013gba,Fox:2013pia} now face increased tension with the LUX results.  Models with exothermic scattering \cite{Graham:2010ca,Frandsen:2013cna,Fox:2013pia,Frandsen:2014ima} are now also in considerable tension with the LUX results, however there is no tension between LUX and an interpretation of the CDMS-Si excess in terms of a DM sub-component such as exothermic double-disk dark matter \cite{McCullough:2013jma}.}

A lesson learned from studying these past DM hints is that the interplay between signals and constraints at different detectors may depend heavily on the local velocity distribution of DM, making this unknown a particularly troubling (or in some cases useful) nuisance parameter \cite{Fairbairn:2008gz,MarchRussell:2008dy,McCabe:2010zh,Peter:2011eu}.  To mitigate this uncertainty, methods which allow the comparison of scattering rates at different detectors irrespective of the DM velocity distribution were developed \cite{Fox:2010bz,Fox:2010bu} and subsequently extended to treat detectors with multiple target nuclei \cite{Frandsen:2011gi}, detector energy resolution effects \cite{Gondolo:2012rs}, annual modulation signals \cite{HerreroGarcia:2012fu}, and inelastic DM scattering \cite{Bozorgnia:2013hsa}\footnote{For a review of halo-independent and related approaches see \cite{Peter:2013aha}.}.  While these methods have been of great utility in scrutinizing past DM hints, they will also be extremely important if a true signal of DM scattering begins to emerge, since the initial stages of discovery would begin with a small statistical excess within a particular detector.  The ability to  compare this to limits from other detectors in a fashion independent of the DM velocity distribution will be critical in assessing the validity of a signal.

While the halo-independent methods are very effective in interpreting null results from DM searches in order to place unambiguous limits on the allowed scattering rates at other detectors, the interpretation of an emerging DM signal using current halo-independent methods is open to some ambiguities.  The current methods require that candidate DM scattering events be grouped into bins of recoil energy.  The total rate in each bin is then mapped into a halo-independent rate to be compared with the limits from other detectors.  For many applications this method is appropriate, however for an emerging DM signal it is not ideal for the following reasons.
\begin{itemize}
\item  State-of-the-art detectors achieve expected backgrounds which are very low, typically expecting $\mathcal{O}(\lesssim\!\! 1)$ background events in the DM acceptance region.  As each new experimental run often leads to less than an order of magnitude increase in sensitivity, an emerging DM signal will likely come in the form of a small number of events.  Many more events may follow with further experimental runs, but it is unlikely that the discovery of DM will begin with a large number of events.  The binning of a small number of events is undesirable, since it is ambiguous and introduces sensitivity to the choice of bins.  Hence, methods which rely on binning will not be optimal in the early stages of DM discovery.
\item  Current and future DM direct detection technology typically achieves excellent energy resolution.  As the uncertainty in the energy of each candidate DM scattering event is likely to be small, bins wider than the energy resolution can only lead to the loss of important information about each event, effectively reducing the interpreted resolution and efficacy of the detector.  Ideally, as much information as possible about each event should be retained in any comparison between candidate DM events and constraints from other detectors.  For an emerging discovery, halo-independent methods which do not rely on binning are desirable.
\end{itemize}

In this work a new halo-independent method for analyzing candidate DM events is proposed which, by the above arguments, would be useful in the early stages of a DM discovery, and beyond.  This builds on previous methods and relies on well-known properties of the integral over the velocity distribution of DM.  The method allows for candidate DM events to be interpreted as best-fit points, with associated confidence intervals, for the DM velocity integral.  These best-fit points and confidence intervals are shown to hold over all possible DM halos, and are in this sense halo-independent.  Once determined, the implied values of the DM velocity integral can then be compared to limits from other detectors, allowing a halo-independent comparison between candidate DM signals and null DM experiments, free from the need to bin events and the ambiguities this introduces.

In \Sec{sec:methods} the halo-independent methods are reviewed.  The calculation of constraints from null experiments is reviewed in \Sec{sec:const} and the new method for an unbinned halo-independent interpretation of candidate DM events is developed and clarified in \Sec{sec:method}.  Comparisons between positive signals and null results are discussed in \Sec{sec:compnull}.  In \Sec{sec:massvary}, we describe how the halo-independent information for one specific DM mass may be simply and unambiguously mapped to other DM masses, avoiding the proliferation of limit plots and calculations.  The reader only interested in a short explanation of how to apply the methods can proceed directly to \Sec{sec:nutshell} where all necessary calculation steps for setting limits and for interpreting signals are briefly set out. In \Sec{sec:application} the new unbinned halo-independent methods are applied to the three anomalous events observed in the CDMS-Si detector and compared to the current constraints from XENON10 and LUX.  Finally, in \Sec{sec:conclusion} conclusions and suggestions for areas of future development are presented.   \App{app:Smearing} contains a proof that our method works equally well for both the idealized case of perfect energy resolution and the more realistic case of finite experimental energy resolution. 

\section{Halo-Independent Analysis Methods}
\label{sec:methods}

The differential event rate\footnote{Throughout this paper we consider only spin-independent coupling of DM to nuclei, the generalization of these techniques to the spin-dependent case is straightforward.} at a direct detection experiment is 
\be
\frac{dR}{d E_R} = \frac{N_A \rho_\chi \sigma_n m_n}{2 m_\chi \mu_{n\chi}^2} C_T^2 (A,Z) \int d E'_R G (E_R,E'_R) \epsilon(E'_R) F^2 (E'_R) g(v_{min} (E'_R)) ~~,
\label{eq:ratea}
\ee
where $m_\chi$ is the DM mass, $m_n$ the nucleon mass, $\mu_{n\chi}$ the nucleon-DM reduced mass, $\sigma_n$ the DM-nucleon scattering cross-section, $\rho_\chi$ the local density, $N_A$ is Avogadro's number, $F (E_R)$ is the nuclear form factor which accounts for loss of coherence as the DM resolves sub-nuclear distance scales, $C_T(A,Z)= (f_p/f_n Z + (A-Z))$ is the usual coherent DM-nucleus coupling factor, $\epsilon (E_R)$ is the detector efficiency, and $G(E_R,E'_R)$ is the detector resolution function. The velocity integral is
\be
g(v_{min}) = \int^\infty_{v_{min}} \frac{f(\mb{v}+\mb{v}_E)}{v} d^3 v ~~,
\ee
where $f(\mb{v})$ is the DM velocity distribution, and $\mb{v}_E$ is the Earth's velocity, both in the galactic frame.  We ignore the small time-dependence introduced by the Earth's motion around the Sun.  For elastically scattering DM the minimum DM velocity required to produce a nuclear recoil energy $E_R$ is
\be
v_{min} (E_R) = \sqrt{\frac{m_N E_R}{2 \mu_{N\chi}^2}} ~~,
\label{eq:vmin}
\ee
where $\mu_{N\chi}$ is the nucleus-DM reduced mass.  As is now standard, the constant factors which are common to all DM detectors are absorbed into a rescaled velocity integral
\be
\tilde{g} (v_{min}) = \frac{\rho_\chi \sigma_n}{m_\chi} g(v_{min})  ~~.
\label{eq:tildeg}
\ee
An observation critical to the halo-independent methods, first noted in \cite{Fox:2010bz,Fox:2010bu}, is that because the velocity integrand is positive definite, $\tilde{g} (v_{min})$ is a monotonically decreasing function of $v_{min}$ for \emph{any} DM halo.  This observation becomes very powerful in developing halo-independent methods for the comparison of multiple experiments, as now described.

\subsection{Constraining $\tilde{g} (v_{min})$ -- null results}
\label{sec:const}
Before considering the possibility of positive DM search results it is worthwhile to first consider the case of null experiments which can be used to constrain the velocity integral $\tilde{g} (v_{min})$.  We follow the discussion of \cite{Fox:2010bz}.  Once a specific value of the DM mass $m_\chi$ is chosen it is possible to place limits on the velocity integral $\tilde{g} (v_{min})$.  If, at some reference minimum velocity $v_{ref}$, the velocity integral is non-zero $\tilde{g} (v_{ref}) \neq 0$ then, since the velocity integral is monotonically decreasing, the unique form for the velocity integral which minimizes the total number of events for a given $\tilde{g} (v_{ref}) \neq 0$ is
\be
\tilde{g} (v_{min}) = \tilde{g} (v_{ref}) \Theta(v_{ref}-v_{min}).
\label{eq:const}
\ee
Thus, for a given choice of DM mass, it is possible to constrain the largest value of $\tilde{g} (v_{ref})$ allowed by a given null experiment by constraining the velocity integral \Eq{eq:const} with standard methods.  As this choice minimizes the total number of events for a given $\tilde{g} (v_{ref})$, limits calculated in this way represent the most conservative limits possible over all halos.  In other words, if a certain value of $\tilde{g} (v_{ref})$ constrained in this way is excluded it is excluded for all possible halos.  However, if it is not excluded by this approach it may still be excluded for many reasonable halos, \eg\ the standard halo model (SHM), but just not for the distribution of \Eq{eq:const} which corresponds to a DM stream at speed $v_{ref}$.  This process is repeated for different values of $v_{ref}$ to build up a continuous exclusion contour in $\tilde{g} (v_{ref})$ over all $v_{ref}$.

\subsection{Discovering $\tilde{g} (v_{min})$ -- positive results}
\label{sec:method}
The most sensitive, and arguably least ambiguous DM detectors, strive to keep backgrounds low enough that $\lesssim \mathcal{O}(1)$ background events are expected in a given run.  They also typically have excellent energy resolution, such that $\Delta E_R/E_R \ll 1$.  These factors combined suggest that the initial emergence of a DM discovery will likely be in the form of a relatively small number, $N_O$, of events observed at discrete energies $E_i$.  To establish the consistency of such a scenario it will be important to then compare this potential DM discovery with limits from other experiments, ideally in a context free of uncertainties in the DM halo.  Clearly an optimal route is to compare constraints on $\tilde{g} (v_{min})$ from the null experiments (described in \Sec{sec:const}) with the non-zero values of $\tilde{g} (v_{min})$ hinted at by the emerging DM signal.

For positive signals, all current methods require the ad-hoc choice of a set of energy bins and then the calculation of upper and lower limits on the signal within these bins using the observed events and estimated backgrounds.  These energy bins and preferred rates in each bin are then converted into $v_{min}$-space bins and preferred values of $\tilde{g} (v_{min})$ in each bin, subject to the constraint that the velocity integral is monotonically decreasing.  The problems with such a method are immediately apparent.  For emerging hints the number of events in the energy range of the detector will be small and binning a small number of events is a statistically questionable exercise from the outset, open to ambiguities and introducing issues with bin choice.  Also, if a detector has good energy resolution, then valuable information is lost by binning the data in bins much larger than the experimental resolution, reducing the efficacy of any interpretation of the DM hint.  Most crucially, binning data in bins of width much greater than the experimental resolution may lead to misinterpretation of the halo-independent constraints on this DM hint.  Conversely, choosing bins of width much smaller than the energy resolution would, in the limit of a small number of events, smear single events across bins.

Ideally, it would be possible to map an emerging DM hint to $\tilde{g} (v_{min})-v_{min}$ space in a way which preserves as much information as possible.  In the case of detectors with excellent energy resolution this is of the utmost importance. But even for detectors with poor energy resolution there is information in the positions of the events and maintaining that information means employing methods which avoid binning the data.  

\subsubsection*{The Method}
A method commonly used in fitting a model with free parameters to unbinned data is the extended maximum likelihood method \cite{Barlow:1990vc} which is desirable over the standard likelihood method as the normalization of a given rate is taken into account.  When applied to a DM direct detection experiment which has observed $N_{O}$ events, in the energy range $[E_{min},E_{max}]$, the extended likelihood is
\be
\mathcal{L} = \frac{e^{-N_{E}}}{N_{O}!} \prod^{N_{O}}_{i=1} \frac{dR_T}{d E_R} \bigg|_{E_R=E_i}~~,
\ee
where $dR_T/dE_R$ contains signal and background components and 
\be
N_{E} = \int_{E_{min}}^{E_{max}}  \frac{dR_T}{d E_R} d E_R ~~,
\ee
is the total number of events expected for a given set of parameters.  We may compare different parameter choices by considering the log-likelihood, $L=-2 \log (\mathcal{L})$ which is minimized for a good fit and grows with decreasing quality of fit.  Discarding constants irrelevant to the fitting procedure we have
\be
L/2 = N_E-\sum^{N_{O}}_{i=1} \log \frac{dR_T}{d E_R} \bigg|_{E_R=E_i}  ~~.
\label{eq:L}
\ee
Using the DM rate, in terms of $\tilde{g} (v_{min})$, as presented in \Eq{eq:ratea}, and including a background component
\begin{eqnarray}
\frac{dR_T}{d E_R}  & = &   \frac{dR_{BG}}{d E_R} + \frac{dR_{DM}}{d E_R} \\
 &= & \frac{dR_{BG}}{d E_R} + \frac{N_A m_n}{2  \mu_{n\chi}^2} C_T^2 (A,Z) \int d E'_R G (E_R,E'_R) \epsilon(E'_R) F^2 (E'_R) \tilde{g}(v_{min} (E'_R)) 
\label{eq:rate}
\end{eqnarray}
where the first term accounts for the (small) estimated backgrounds and the last term the DM signal.  There now appears to be a barrier to calculating $L$ since there are an infinite set of possible DM halos to consider as one must also make a choice of the form of $\tilde{g}(v_{min} (E_R))$ not only at each event, but over the whole range of measurable energies since the total number of events is calculated as the integral over this energy range.

\begin{figure}[t]
\centering
\includegraphics[height=2.65in]{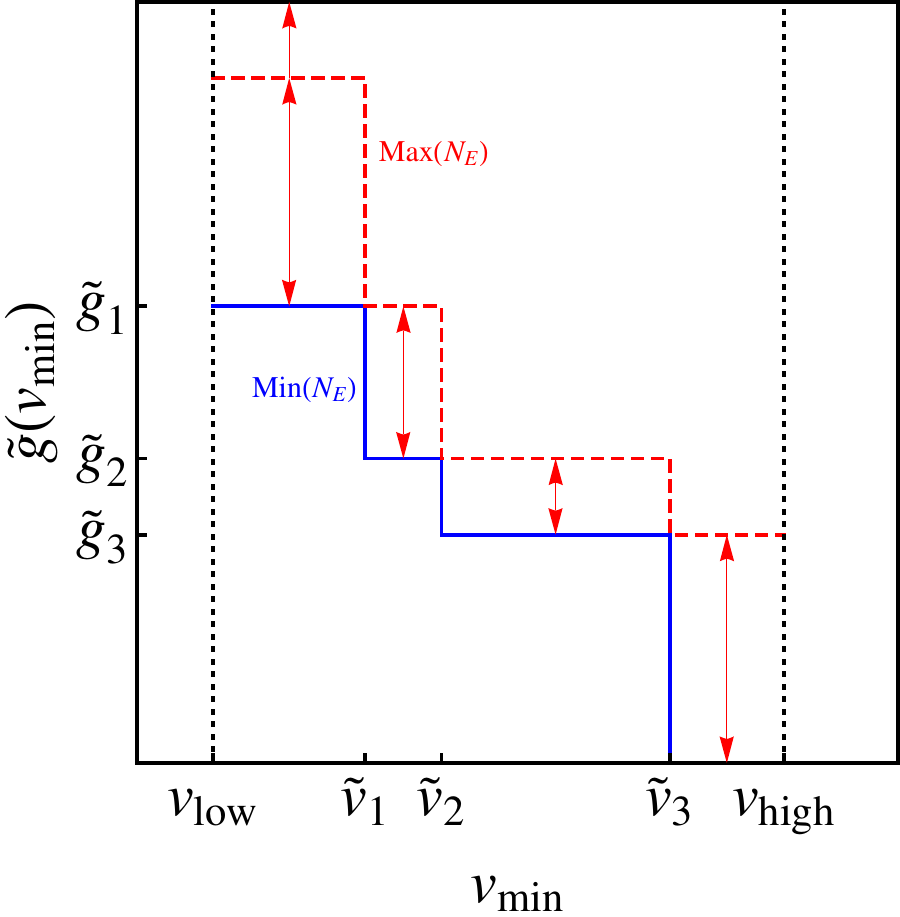}
\caption{A schematic representation of all halo possibilities for $\tilde{g} (v_{min})$.  If an experiment observes a number of events consistent with DM scattering, in this case three events of energy $E_i$, then hypothetical values of $\tilde{g}(\tilde{v}_{i-1} < v_{min} \leq \tilde{v}_i) = \tilde{g}_i$ may be chosen where the positions of the steps $\tilde{v}_i$ are given by $v_{min} (E_i)$ in the case of perfect energy resolution, and are allowed to float as free parameters if the energy resolution is non-zero.  The solid blue curve will always minimize the extended log-likelihood, both in the case of perfect energy resolution and also with resolution effects included as demonstrated in \App{app:Smearing}.  Conversely the dashed red curve corresponds to the worst possible fit out of all halos, which is infinitely bad if the velocity integral between $v_{low}$ and $v_1$ is taken to infinity.  Here, $v_{low}$ ($v_{high}$) is the velocity that corresponds to the low (high) energy threshold of the experiment.  To determine the range of halos implied by the DM candidate events the parameters $\tilde{g}_i$ and $\tilde{v}_i$ may be varied, consistently choosing the solid blue curve in the likelihood, in order to determine the best-fit values and confidence intervals for $\tilde{g}_i$.}
\label{fig:AllHalo}
\end{figure}

For simplicity let us first consider the case with perfect energy resolution $G (E_R,E'_R) = \delta (E_R-E'_R)$.  A given set of events corresponds to a set of $N_O$ hypothetical values of $\tilde{g}_i \equiv \tilde{g}(v_{min}(E_i))$ as well as the form of $\tilde{g}(v_{min} (E_R))$ interpolating between the $\tilde{g}_i$.  However, \Eq{eq:L} penalizes against the total number of events predicted, since $L$ increases as $N_E$ increases.  Thus, since $\tilde{g}(v_{min} (E_R))$ is monotonically decreasing, the \emph{best} fit out of all possible DM halos is the one which minimizes the total number of events predicted in any interval $E_{i-1} < E_R < E_{i}$ between events. This is accomplished by choosing a constant value $\tilde{g}(v_{min} (E_{i-1}<E_R\leq E_{i})) = \tilde{g}_i$,\footnote{We define $E_0$ to be the lower threshold of the experiment, $E_{min}$.} which is illustrated in \Fig{fig:AllHalo}. 

This form of $\tilde{g}(v_{min})$ is quite robust. Indeed, in \App{app:Smearing} we prove using variational techniques that the best-fit $\tilde{g}(v_{min})$ is still a sum of $N_O$ step functions even in the case of a very general resolution function; the only difference is that the positions  $\tilde{v}_i$ of the steps may now shift to the right of their position in the scenario with perfect energy resolution, $\widetilde{v}_{i}\geq v_{min}(E_{i})$. Thus, in all cases of interest, the form of the velocity integral which minimizes the extended likelihood for $N_O$ observed events is a sum of at most $N_O$ step functions,\footnote{Two step functions of the same height are equivalent to one step function, so in practice there may be fewer than $N_O$ steps.} whose $2N_O$ free parameters (heights and positions) may be determined numerically in a straightforward manner, or analytically in the case of perfect energy resolution.

To calculate the log-likelihood it helps to define
\be
\mu_i = \frac{dR_{BG}}{d E_R} \bigg |_{E_i} ~~,
\label{eq:diffbg}
\ee
the differential background rate evaluated at the energy of each event $E_i$, and
\be
\tilde{\mu}_i = \frac{dR_{DM}}{d E_R} \bigg |_{E_i} = \frac{N_A m_n}{2  \mu_{n\chi}^2} C_T^2 (A,Z) \sum_{j=1}^{N_O} \tilde{g}_j \int_{\widetilde{E}_{j-1}}^{\widetilde{E}_{j}} d E'_R G (E_i,E'_R) \epsilon(E'_R) F^2 (E'_R)  ~~,
\label{eq:diifunit}
\ee
which is the differential scattering rate at each event $E_i$. Here $\widetilde{E}_{i}$ are the positions of the steps in the halo velocity integral $\tilde{g}$ (written as a function of recoil energy $E_R$) satisfying $\widetilde{E}_{i}=E_i$ in the case of perfect energy resolution. Another useful quantity is
\be
\widetilde{N}_{T} =  \frac{N_A m_n}{2  \mu_{n\chi}^2} C_T^2 (A,Z) \sum_{j=1}^{N_O} \tilde{g}_j \int_{\widetilde{E}_{j-1}}^{\widetilde{E}_{j}} d E'_R \epsilon(E'_R) F^2 (E'_R)  ~~,
\label{eq:np}
\ee
which is simply the total number of DM events expected.\footnote{The resolution function has already been integrated over in this expression.}  In terms of these quantities (which depend on the $\tilde{g}_i$ and the $\tilde{E}_j$) the extended log-likelihood now decomposes as,
\begin{eqnarray}
L & = & \sum^{N_{O}}_{i=1} L_i = 2 \left(  \widetilde{N}_{T} + N_{BG} -\sum^{N_{O}}_{i=1} \left( \log{\left( \tilde{\mu}_i + \mu_i \right)} \right)\right) \\
& \rightarrow &  2 \left(  \widetilde{N}_{T} - \sum^{N_{O}}_{i=1} \left( \log{\left(\tilde{\mu}_i + \mu_i \right)} \right)\right)~,
\label{eq:likeall}
\end{eqnarray}
where in going to the last line irrelevant constants have again been discarded.  In this way the construction of the likelihood function for $N_O$ events simply requires the straightforward calculation of the quantities defined in \Eq{eq:diffbg}, \Eq{eq:diifunit}, and \Eq{eq:np}.

\Eq{eq:likeall} contains all of the information required to find the best-fit values and confidence intervals for the DM halo integral. To find the best-fit values $\tilde{g}_{i,min}$ and the best-fit positions of the steps $\widetilde{E}_{i,min}$, the likelihood may be numerically minimized to find $L_{min}$, subject to the monotonicity constraint which must be imposed for any DM interpretation \ie\ $\tilde{g}_{i,min} \geq \tilde{g}_{i+1,min}$.  The confidence intervals in each $\tilde{g}_i$, denoted $\Delta \tilde{g}^{\pm}_i$, may be found by determining the extremum values satisfying $L(\tilde{g}_i\pm\Delta \tilde{g}^{\pm}_i) = L_{min}+\Delta L$, for some $\Delta L$ which is determined from the statistical confidence desired.  The other values of $\tilde{g}_{j\neq i}$ and the positions of the steps $\widetilde{E}_{j}$ should also be allowed to vary when determining the extremum values.  It should be noted that in determining the confidence intervals, the monotonicity constraint must still be imposed, thus in determining $\Delta \tilde{g}^{\pm}_i$ the other $\tilde{g}_{i'\neq i}$ may not always take their best-fit values.

To determine the allowed region at a given statistical confidence level, one would typically use the $\Delta L$ corresponding to the $\chi^2$ value for the number of parameters in the fit.  However, this approach breaks down when the number of events is small. Furthermore, since parameter points which extremize $\Delta L$ typically live on the boundary of the parameter space, where the monotonicity constraint is saturated (possibly multiple times), the constraint reduces the number of effective parameters.  Thus the determination of $\Delta L$ is best done through Monte Carlo simulation.  Taking the underlying probability distribution function to be given by the best fit values $(\widetilde{E}_{min},\tilde{g}_{min})_i$, combined with the background model, one generates a large set of fake data.  Each iteration contains the same number of events as was observed in the experiment.  For every pseudo-experiment the likelihood is extremized as before and the best fit values for that run are recorded.  For a large enough set of pseudo-experiments, the mean best fit values $(\overline{E},\overline{\tilde{g}})_i$ should lie close to the original best fit found for the actual data.  Together the mean $\vec{\mu}$ and the covariance matrix $\sigma$ of the best-fit parameters define the surface of a hyper-ellipsoid of radius $\sqrt{\Delta L}$ in parameter space,
\begin{equation}
(\vec{x}-\vec{\mu})\sigma^{-1}(\vec{x}-\vec{\mu})=\Delta L.
\end{equation}
The $\Delta L$ corresponding to, for example, 90\% C.L. is determined by the radius of the hyper-ellipsoid that contains 90\% of the pseudo-experiments.  The region of parameter space that contains the best-fit parameters for the actual data at 90\% confidence is within this $\Delta L$ of the actual best fit $L_{min}$.

As an infinite number of possible halos have been discarded, one may wonder whether this method actually captures the full ranges for $\tilde{g}_i$ at the desired confidence level.  For a given $\tilde{g}_i$, a non-minimal halo not saturating the monotonicity constraint, \ie\ one for which $\tilde{g} \left( v_{min} (E_{i-1}<E_R<E_{i}) \right) > \tilde{g}_i$, would only increase the value of $\widetilde{N}_{T}$ and therefore the log-likelihood, meaning that a \emph{smaller} range of $\tilde{g}_i$ would be allowed with respect to the global minimum of the likelihood.  Thus, rather than testing all possible halos to determine the best-fit values of, and allowed range of, the $\tilde{g}_i$ one can instead make the the minimal (saturating) choice, $\tilde{g}(v_{min} (E_{i-1}<E_R < E_{i})) = \tilde{g}_i$.  The best-fit points found this way will be the best fit out of \emph{all} possible halos and the confidence intervals $\Delta \tilde{g}_i^\pm$ necessarily encompass the maximally allowed ranges.  This means that the envelope of allowed $\tilde{g}_i$ captures all halos for which the extended likelihood is within $\Delta L$ of the minimum.

\subsection{Comparing with null results}
\label{sec:compnull}
Although a DM hint may suggest non-zero values of $\tilde{g}_i$ for each anomalous event, it is desirable to compare these values in a halo-independent way with constraints from detectors which do not observe a signal.  As described in \cite{Fox:2010bz}, and \Sec{sec:const}, the most conservative limit on the velocity integral at a specific value of $v_{min}=v_{ref}$, denoted $\tilde{g} (v_{ref})$, may be determined by considering limits on the function $\tilde{g} (v_{min}) = \tilde{g} (v_{ref}) \Theta(v_{ref}-v_{min})$.

Calculating limits on $\tilde{g} (v_{min})$ in this way, and the best-fit values and confidence intervals for $\tilde{g}_i$ suggested by a DM hint using the method above, leads to plots such as \Fig{fig:CDMSLUX}, showing experimental limits and the best-fit values and confidence envelope for the velocity integral.  It should be emphasized that the envelope of $\tilde{g} (v_{min})$ does not imply that \emph{any} curve passing through the envelope will have a log-likelihood value of $L \leq L_{min}+\Delta L$, but it does imply that there exists a curve which passes through any single point in the envelope within a confidence interval satisfying $L \leq L_{min}+\Delta L$.  Furthermore, no curve with $L \leq L_{min}+\Delta L$ lies outside the envelope.

The most important information in any such plot is the interplay between the limits curve and the preferred envelope in the velocity integral.  Consider a point on the lowest boundary of the envelope in $\tilde{g} (v_{min})$ at a point $v'_{min}$, denoted $\tilde{g}_{-} (v'_{min})$.  The halo which leads to this value of $\tilde{g} (v'_{min})$ at $v'_{min}$, but would predict the smallest possible number of events in any detector, corresponds to $\tilde{g} (v_{min}) = \tilde{g}_{-} (v'_{min}) \Theta(v'_{min}-v_{min})$.  However, it is precisely this halo shape which has been constrained by the null experiment.  Hence, if a single point along the lowest boundary of the preferred envelope for $\tilde{g} (v_{min})$ is excluded by a null experiment then there is no halo within $\Delta L$ of the minimum of the likelihood for which the hint could be consistent with the null experiment.  In other words, it is excluded by the null experiment independent of any uncertainties in the DM halo.

\subsection{Varying $m_\chi$}
\label{sec:massvary}
The halo-independent methods are clearly of great value in comparing experimental results whilst avoiding the significant uncertainties in the velocity distribution of the DM.  One perceived weakness of this approach is that it appears the calculations must be performed under the hypothesis of a single DM mass, $m_\chi$, and to consider a different DM mass $m_\chi'$ the entire calculation must be repeated again, leading to a proliferation of plots when presenting the results.  However, assuming the detector is built from a single material, once limits and best-fit velocity integrals have been calculated for a single DM mass $m_\chi$, it is simple to map them to the analogous quantities for a different mass $m_\chi'$. 

Let us first consider the energy of a scattering event.  The minimum DM velocity required is given by \Eq{eq:vmin} which, for a specific scattering energy, immediately gives the relationship between $v_{min} (E_R)$ for a DM mass $m_\chi$ and $v'_{min} (E_R)$ for a DM mass $m_\chi'$,
\be
v'_{min} (E_R) = \frac{\mu_{N\chi}}{\mu_{N\chi'}} v_{min} (E_R) ~~,
\label{eq:mapvmin}
\ee
mapping a point on the $v_{min}$ axes for $m_\chi$ to a point on the $v'_{min}$ axes for $m'_\chi$ while preserving the ordering of the scattering events.  It should be noted that this mapping is nucleus-dependent, and shifts limits and hints from different detectors by differing amounts.  Furthermore, as halo-independent limits and best-fit points are calculated assuming a flat velocity integral between any neighboring events, the total number of events predicted between any two events only changes by a global normalization factor.  This normalization can be found from \Eqs{eq:ratea}{eq:tildeg}, where it is clear that under a change in the DM mass, $m_\chi \to m_\chi'$, the required normalization of $\tilde{g}$, whether as a best-fit point, or a point on an exclusion curve, will be shifted to
\be
\tilde{g}' = \frac{\mu^2_{n\chi'}}{\mu^2_{n\chi}} \tilde{g} ~~.
\label{eq:maptildeg}
\ee
These two transformations, \Eq{eq:mapvmin} and \Eq{eq:maptildeg}, define a unique mapping between any point on the $\tilde{g}-v_{min}$ plane for a DM mass $m_\chi$, to a new point on the $\tilde{g}'-v'_{min}$ plane for a DM mass $m_\chi'$.

This has important implications for the presentation of DM direct detection results: if new DM limits, or hints, are presented in plot form in the halo-independent framework for a specific DM mass, then a single plot alone contains all of the information required \emph{for all DM masses}.  Thus, if an experimental collaboration released such a plot it would be possible to study halo-independent limits for any DM mass.  Even if many details of the experimental analysis are not publicly available, this would enable the robust application of the DM results to different halo-independent scenarios by external groups.

As the shift in the normalization affects all $\tilde{g}$ equally, the same minimum value of the log-likelihood \Eq{eq:likeall} will be found for \emph{any} DM mass, and the halo-independent method contains no information on the preference of data from a single experiment for a specific DM mass.  A preferred mass may only be determined by appealing to a specific halo, or through requirements on the upper limits on $v_{min}$ due to the galactic escape velocity, or by combining date from multiple experiments.

\subsection{In a Nutshell}
\label{sec:nutshell}

It is useful at this point to summarize the steps required to perform an unbinned halo-independent analysis with real data.  To calculate exclusion contours in $\tilde{g} (v_{min})$ from a null experiment, it is only necessary to calculate limits in the usual way, with the exception that at a  point $v_{ref}$ the usual velocity integral is replaced with $\tilde{g} (v_{min}) = \tilde{g} (v_{ref}) \Theta(v_{ref}-v_{min})$ and an upper bound is calculated for the constant $\tilde{g} (v_{ref})$.  This process, which was described in \cite{Fox:2010bz} is repeated for different values of $v_{ref}$ to build up an exclusion contour.

In the case of an experiment with good energy resolution which observes an excess of events over an expected $\mathcal{O}(\lesssim 1)$ background events, it is only necessary to assume the velocity integral $\tilde{g} (v_{min})$ takes the form of at most $N_O$ step functions with undetermined heights and positions.  It is then necessary to calculate $\mu_i$, $\tilde{\mu}_i$, for each event and the total number of events predicted $\widetilde{N}_{T}$, where these quantities are defined in \Eq{eq:diffbg}, \Eq{eq:diifunit}, and \Eq{eq:np}.  With these quantities in hand one simply varies the heights and positions of the steps to find the minimum of the sum $L/2= \widetilde{N}_{T} - \sum^{N_{O}}_{i=1} \log{\left(\tilde{\mu}_i + \mu_i \right)}$ with the additional constraints that $\tilde{g}_{i,min} \geq \tilde{g}_{i+1,min}$.  The uncertainty on these determinations, at a given confidence level, is given by finding the variations, $\pm\Delta \tilde{g}_{v_{min}}$, which saturate $L=L_{min}+\Delta L$ (also allowing the positions of the steps to vary) to construct an envelope of preferred values for $\tilde{g}(v_{min})$.  The ranges $\pm\Delta \tilde{g}_i$ encapsulate the full envelope of possibilities of all DM distributions which are monotonic and satisfy $L=L_{min}+\Delta L$.  The determination of the relevant $\Delta L$ is best done by carrying out pseudo-experiments, as described in section \ref{sec:method}.

Once these limits and hints have been calculated and compared for a particular DM mass they have effectively been compared for all masses, assuming the DM scatters elastically and the detector consists of a single target.

\section{Application to real data: CDMS-Si versus XENON and LUX}
\label{sec:application}
The new halo-independent method is now employed to investigate the consistency between the $\sim 3 \sigma$ excess of events observed by the CDMS-Si collaboration \cite{Agnese:2013rvf} and the most constraining null results from the xenon-based detectors, which are currently the XENON10 and LUX experiments.  This not only illustrates the utility of the method for detectors with good energy resolution and a small number of observed events, but also represents the first halo-independent unbinned comparison between the CDMS-Si excess and the recent LUX results.

The S2-only XENON10 analysis \cite{Angle:2011th} is used, with the ionization yield $\mathcal{Q}_y$ also taken from \cite{Angle:2011th}.  We take the detector resolution function $G (E_R,E'_R)$ to be a Gaussian with energy-dependent width $\Delta E_R = E_R / \sqrt{E_R \mathcal{Q}_y (E_R)}$. The acceptance is $95\%$, and the exposure is $15$ kg days.  Yellin's `Pmax' method \cite{Yellin:2002xd} is used to set limits.

\begin{figure}[t]
\centering
\includegraphics[height=2.5in]{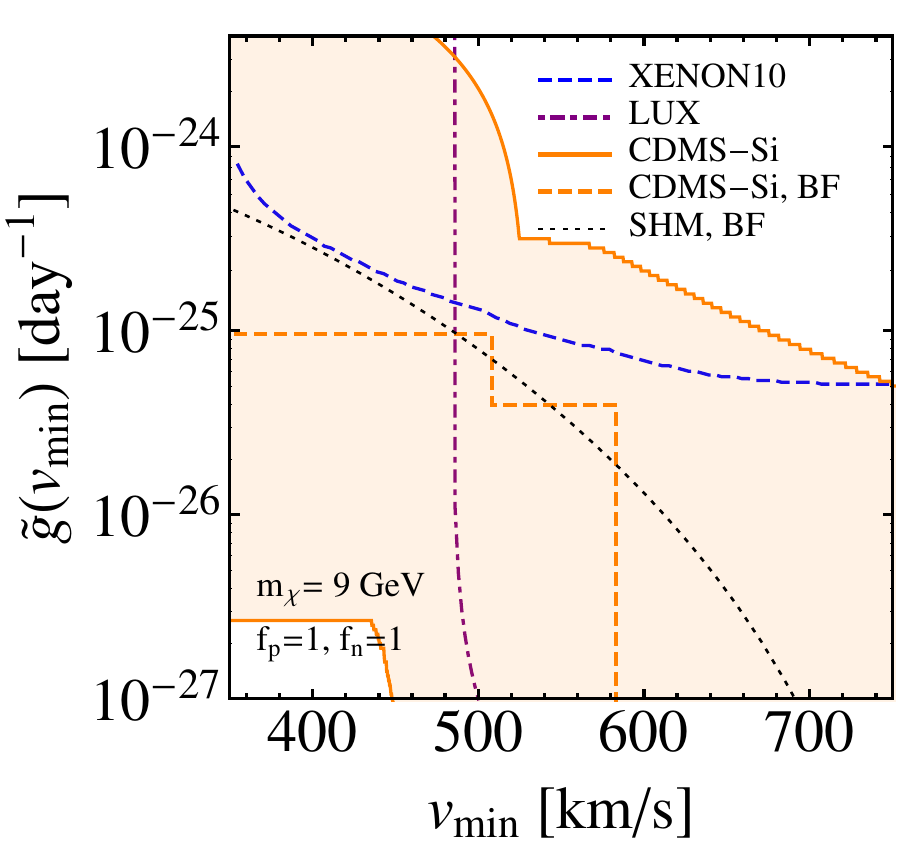} \hspace{0.2in} \includegraphics[height=2.5in]{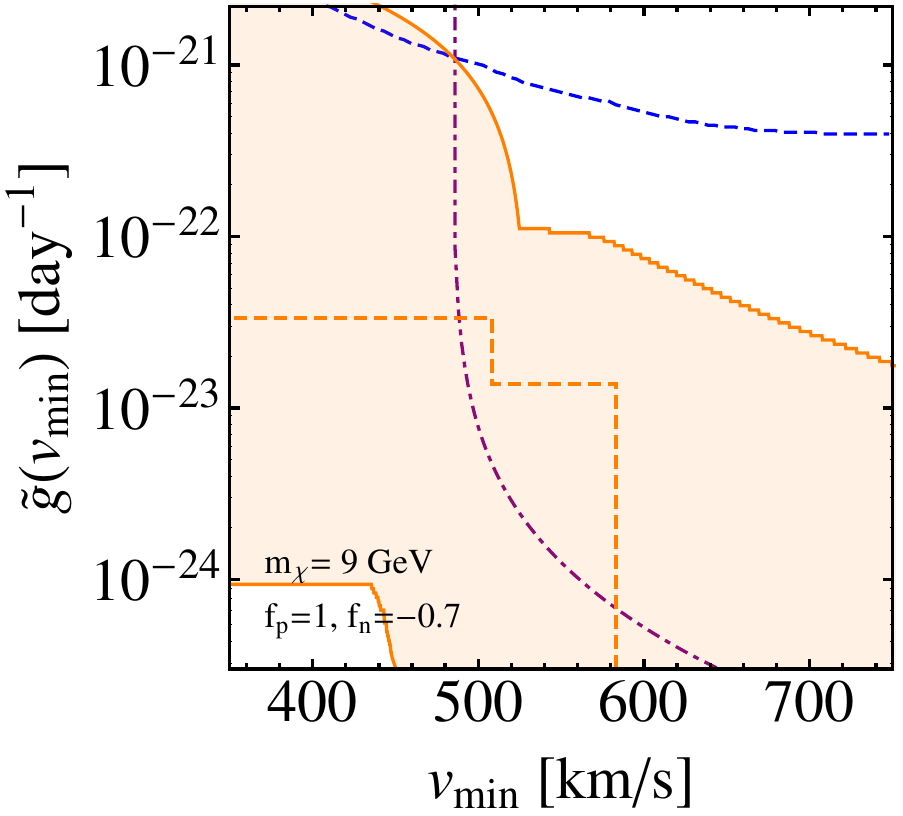}
\caption{Halo-independent interpretation of the CDMS-Si events versus constraints from XENON10 and LUX assuming elastic, spin-independent scattering with equal couplings to protons and neutrons (left panel) and with couplings tuned to maximally suppress the sensitivity of xenon experiments (right panel).  The preferred envelope and constraints are both calculated at $90\%$.  The best-fit halo is inconsistent with the LUX results and only a small section of the lower boundary of the preferred halo envelope for CDMS-Si is compatible with the null LUX results, meaning that only a small range of DM halos are compatible with the LUX results for which the extended likelihood is within $\Delta L$ of the best-fit halo.  If the DM-nucleon couplings are tuned to maximally suppress scattering on xenon, the best-fit DM interpretation is still inconsistent with the LUX results, however the range of viable halos is increased.  The curve for the SHM is also shown, giving a good fit to the CDMS-Si data as well as a curve for the best-fit halo which minimizes the extended likelihood.}
\label{fig:CDMSLUX}
\end{figure}

The LUX collaboration have recently announced results from the first run \cite{Akerib:2013tjd}.  The estimated LUX background distributions are not yet publicly available, making a profile likelihood ratio (PLR) test statistic analysis impossible.  In \cite{LUXtalk} it was shown that for light DM the vast majority of nuclear recoil events would actually lie below the mean of the AmBe and Cf-252 nuclear recoil calibration band.  The reason for this is that for a given low S2 signal the S1 signal is likely to have appeared above threshold due to a Poisson fluctuation.  As there are no events in the region expected for light DM scattering (or equivalently low energy events) the DM event detection efficiency provided in \cite{Akerib:2013tjd} can be used to calculate the total number of expected events for a light DM candidate and then a Poisson upper limit can be set for zero observed events.  We find excellent agreement with the estimated limits from \cite{DelNobile:2013gba} and good agreement with the official LUX results for the light DM region.

For CDMS-Si three events were found in $140.2$ kg days of data \cite{Agnese:2013rvf}.  We take the detector resolution function $G (E_R,E'_R)$ to be a Gaussian and assume a conservative detector resolution of 0.5 keV. The acceptance is taken from \cite{Agnese:2013rvf}.  The background contributions are taken from \cite{APStalk} with normalization such that surface events, neutrons, and ${}^{206}$Pb, give  $0.41$, $0.13$, and $0.08$ events respectively.  The best-fit points and confidence regions are calculated following the method described in \Sec{sec:method}, the confidence intervals are calculated for a variation $\Delta L=9.2$, where $L$ is the total log-likelihood.  This value of log-likelihood corresponds to a chi-squared distribution for six degrees of freedom and one constraint, thus five free parameters altogether where there are six degrees of freedom from the heights of each step and the step positions and there is one constraint due to the monotonicity constraint.  As parameter points which extremize $\Delta L$ typically live on the boundary of the parameter space where the constraint is saturated the constraint effectively reduces the number of effective parameters.  Thus we choose $\Delta L = 9.2$ as this corresponds to the $\chi^2$ value for five parameters and a confidence interval of $90\%$.  We were led to this choice numerically by generating large sets of fake data from a given underlying three-step-function distribution.  For each set of fake data we then perform the usual procedure of allowing a step for each event, and then varying the heights and positions of the steps to find the best-fit halo for those events.  We then compare the best-fit value of the log-likelihood for these generated events to the best-fit value for the true underlying halo and find that $90\%$ of the results lie within a distribution which we find to be very well approximated by a $\chi^2$.\footnote{We thank Brian Feldstein and Felix Kahlhoefer for conversations regarding the choice of log-likelihood.}

In \Fig{fig:CDMSLUX} we show the halo-independent constraints on an elastically scattering spin-independent DM scattering interpretation of the CDMS-Si events.  There is some tension between the CDMS-Si excess and the LUX results independent of the DM halo if the DM couples equally to protons and neutrons.  The lower energy events may still be consistent with LUX, however with a reduced number of events the significance of the excess is reduced.  Even when couplings are tuned to maximally suppress scattering on xenon \cite{Feng:2013vod}, the best-fit elastically scattering DM interpretation of the highest energy event is in tension with the LUX results.  The best-fit halo interestingly takes the form of two step functions.  Although there are three events, the Gaussian smearing leads to a best-fit halo which only has two steps, whereas in the case of perfect energy resolution there would be three steps. 

Thus, independent of uncertainties in the DM halo, and free from uncertainties introduced by binning the three anomalous CDMS-Si events, a DM interpretation of this excess faces some tension with the LUX results.  This tension is reduced if the DM-proton and neutron couplings are tuned to maximally suppress scattering on xenon, however even when exploiting this freedom there is still tension with the LUX results.  A $\tilde{g} (v_{min})$ curve is also shown for the SHM to demonstrate that the CDMS-Si events give a good fit to the SHM.  We also note that the CDMS data alone prefer a DM contribution over a background-only description, $\tilde{g}\equiv 0$.  This is not surprising, since by allowing for general speed distributions, the lowest energy excess event in any data can always be fit by the DM hypothesis, and thus the overall fit can be improved.  Unlike the relative quality of the fits from the background-only hypothesis compared to signal plus background, the absolute quality of the fit cannot be determined by the methods employed here.  Approaches which determine the goodness of fit but do not requiring binning have been developed (for a review see \cite{Williams:2010vh}), but their application to small data sets is not well understood.  To determine the behaviour of these techniques for the small number of events in direct detection experiments would require extensive modelling in Monte Carlo, which is beyond the scope of this work.

\begin{figure}[t]
\centering
\includegraphics[height=2.5in]{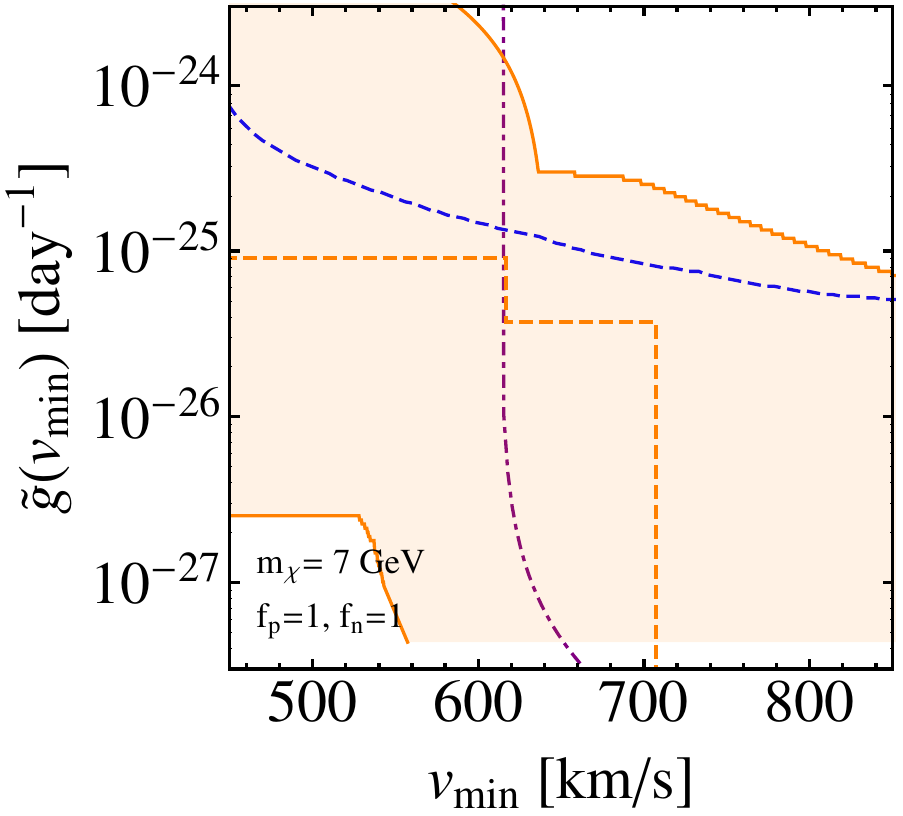} \hspace{0.2in} \includegraphics[height=2.5in]{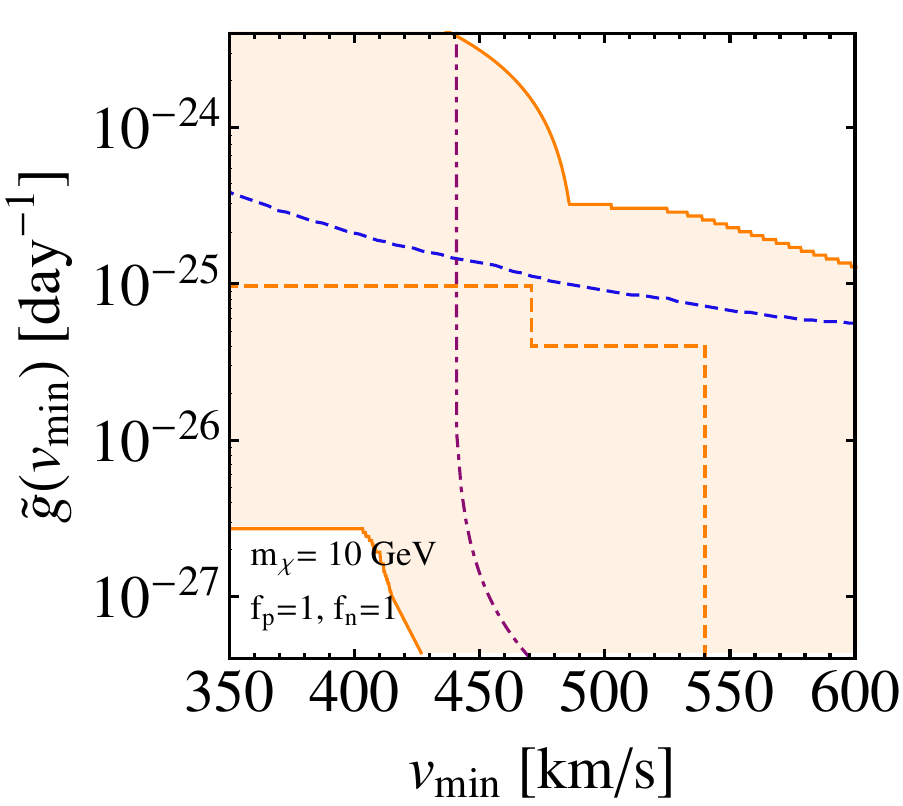}
\caption{The same as \Fig{fig:CDMSLUX} with the mapping of \Eq{eq:mapvmin} and \Eq{eq:maptildeg} employed to calculate the halo-independent limits for $m_\chi=7$ GeV and $m_\chi=10$ GeV directly from the limits for $m_\chi=9$ GeV shown in \Fig{fig:CDMSLUX}.}
\label{fig:CDMSLUXmap}
\end{figure}

In \Fig{fig:CDMSLUXmap} we show the result of using the mapping, \Eq{eq:mapvmin} and \Eq{eq:maptildeg}, from the points in \Fig{fig:CDMSLUX} for $m_\chi=9$ GeV to curves for other masses, demonstrating that an exclusion curve, or best-fit points, for a single DM mass contains all of the information necessary to translate the curve of best-fit points in a halo-independent way for different masses.  This confirms that the presentation of new experimental results in a halo-independent manner for a single DM is a very efficient way to communicate the halo-independent information. 

\section{Conclusions}
\label{sec:conclusion}
The DM direct detection field continues to evolve rapidly. The richness and effectiveness with which this dark frontier is explored relies on multiple experiments and detection strategies being employed.  If an experiment begins to observe events consistent with DM scattering it will be crucial to confirm or refute this possibility with a separate independent experiment which uses different techniques and a different target nucleus.  Previously developed halo-independent methods significantly reduce the systematic errors in such a comparison by eliminating the uncertainties due to the unknown DM velocity distribution.  In this work these methods have been extended to enable a halo-independent analysis of candidate DM events without having to resort to event binning, which is inappropriate for a small number events and for detectors with good energy resolution, as would be expected in the circumstances of an emerging DM discovery.  This method was developed for the simplest scenario of elastically scattering DM, however it would be interesting to extend it to include non-minimal scenarios such as inelastic or exothermic DM, or non-isotropic scattering. 

The method we have described uses the standard approach of minimizing the extended likelihood, which has the advantage of being a well known technique in the field and thus straightforward for experimental collaborations to implement.  Furthermore, it has the feature that results from multiple experiments can be straightforwardly added to the likelihood to carry out a combined analysis, although we have not studied such combinations here. This is true for both excesses and limits.  It would be interesting to see if other statistical techniques, which do not require binning, give similar results. In addition to being straightforward for the experimental collaborations to implement, and reducing one of the systematic uncertainties that plague the interpretation of their results, we reemphasize that this does not come at the expense of complicating the presentation of their results.  For DM scattering elastically in a detector with a single target, the results need only be presented for a single DM mass as this contains all necessary information; the extension to other masses is straightforward to calculate from the results for a single mass.    In addition this method provides a halo-independent analogue of the usual comparison between limits and preferred regions.

Finally, as a test example we applied our technique to the recent results from CDMS and LUX.  In accordance with expectations an unbinned halo-independent analysis of the three anomalous CDMS-Si events shows that for elastic, spin-independent scattering the CMDS-Si events are in tension with the null results from the LUX detector.  If a DM interpretation of the CDMS-Si excess is to be found with no tension from the LUX results, this analysis suggests it will require non-standard DM scenarios.

\section*{Note added}
While this work was in final preparation \Ref{Feldstein:2014gza} was made public.  While also concerned with halo-independent analyses \Ref{Feldstein:2014gza} is based on event binning and is thus complementary to the method proposed here using unbinned methods.

\acknowledgments{We would like to thank Prateek Agrawal, Kyle Cranmer, Brian Feldstein, Felix Kahlhoefer, Joe Lykken, Christopher McCabe, and Jesse Thaler for conversations.  MM thanks the Princeton PCTP workshop ``The Dark Matter Paradigm: Current Status and Challenges'' for stimulating a conversation with David J. E. Marsh which motivated this work.  MM is grateful for the support of a Simons Postdoctoral Fellowship.  YK thanks Grace Haaf, Joshua Batson, and Tiankai Liu for helpful discussions about functional optimization with inequality constraints. YK is supported by an NSF Graduate Research Fellowship. PF and MM would like to thank the Belfast Education \& Library Board for support during the \emph{very} early stages of this work.  Fermilab is operated by Fermi Research Alliance, LLC under Contract No. DE-AC02-07CH11359 with the United States Department of Energy.}

\appendix
\section{Optimal halos and finite energy resolution}
\label{app:Smearing}

For an experiment with finite energy resolution $G(E_R, E_R')$, one may worry that due to smearing effects, the halo integral which minimizes the log-likelihood is no longer a sum of step functions, but perhaps a more complicated function whose many free parameters preclude a simple numerical minimization of the kind described in \Sec{sec:nutshell}. Here we present a proof to the contrary -- for any physically reasonable resolution function, the \emph{only} effects of smearing are to shift slightly the positions of the steps of $\tilde{g}(E_R)$ away from the measured energies $E_i$, and possibly to merge some of the steps. In particular, the optimal halo integral is still a sum of at most $N_O$ step functions.

Although we have in mind Gaussian smearing, this analysis holds for any reasonable form of the resolution function.  We define a physically reasonable resolution function $G(E_R, E_R')$ to have the following properties:
\begin{enumerate}[(i)]
\item $\int G(E_R, E_R') dE'_R = 1$ for any $E_R$.
\item As a function of $E_R'$ for fixed $E_R$, $G(E_R, E_R')$ has a single local maximum at $E_R' = E_R$ and no other local extrema.
\item For $E_R \neq E_R'$, either $G(E_R, E_R') = 0$ or $\partial G(E_R, E_R')/\partial E_R' \neq 0$.
\end{enumerate}
Property (i) simply states that the resolution function is normalized and doesn't change the total number of events. Property (ii) states that $G$ has a single peak where the detected energy equals the true energy, and no other structure. Property (iii) is a technical assumption which will be used in the arguments below, and states that if $G$ is flat on some interval, it must vanish. A normalized Gaussian resolution function $G(E_R, E_R') \propto e^{-(E_R - E_R')^2/2\sigma^2}$ clearly satisfies all three properties,\footnote{A Gaussian with an energy-dependent width $\sigma(E_R)$ also satisfies these properties as long as the form of $\sigma(E_R)$ is physically reasonable. For example $\sigma(E_R) \sim 1/\sqrt{E_R}$ in the XENON experiment, and $G(E_R, E_R')$ satisfies property (ii) as long as the region of $E_R$ close to zero is avoided.} as does a delta function $G(E_R, E_R') = \delta(E_R - E_R')$. Certain models for energy resolution may violate property (iii), for example a ``top hat'' shape where $G(E_R, E_R')$ is constant in some interval about $E_R$ and zero everywhere else, but one may always assume some infinitesimal deviations from flatness which otherwise has no measurable effect.

To simplify the notation, we write the differential scattering rate (\ref{eq:ratea}) as
\be
\frac{dR}{dE_R} = \int dE'_R \, G(E_R, E'_R)K(E'_R)\tilde{g}(E'_R),
\ee
where we have absorbed the form factor, efficiency, and all prefactors into $K(E'_R)$. For reasonable choices of the form factor and efficiency functions, $K(E'_R) > 0$ for all $E'_R$ within the experimental sensitivity, and in addition $dK/dE_R'$ is small. We have also written $\tilde{g}(E'_R)$ as a function of $E'_R$ directly rather than $v_{min}$. Note however that since $v_{min}$ is a monotonic function of $E_R'$, $\tilde{g}(E_R')$ is also monotonic function of $E_R'$. Consider now the expression for the log-likelihood (\ref{eq:L}), written in the suggestive form
\be
L[\tilde{g}] = \int dE'_R \, K(E'_R)\tilde{g}(E'_R) - \sum_{i=1}^{N_O} \log \left ( \mu_i + \int dE'_R \, G(E_i, E'_R)K(E'_R)\tilde{g}(E'_R) \right ).
\label{eq:LLag}
\ee
Property (i) above ensures the resolution function $G$ does not appear in the first integral.

We can now view the log-likelihood minimization as a variational problem: minimize the functional $L[\tilde{g}]$ with respect to the function $\tilde{g}(E_R')$, subject to the monotonicity constraint $d\tilde{g}/dE_R' \leq 0$. The subject of variational problems with inequality constraints may be somewhat unfamiliar to physicists, but is well-known in economics and related fields; the solution is given by the Karush-Kuhn-Tucker conditions \cite{Karush, KuhnTucker}, which generalize the concept of Lagrange multipliers.  In a similar fashion to imposing an equality constraint with a Lagrange multiplier, we can impose the inequality constraint by introducing an auxiliary function $q(E_R')$ and modifying the log-likelihood, $L[\tilde{g}]\rightarrow L[\tilde{g}]+\int dE'_R \, \frac{d\tilde{g}}{d E'_R} q(E'_R)$. The solution that minimizes the log-likelihood while satisfying the monotonicity constraint will satisfy
\begin{align}
\frac{\delta L}{\delta \tilde{g}} - \frac{dq}{dE_R'} & = 0~, \label{eq:var}\\
\frac{d\tilde{g}}{d E_R'} & \leq 0~, \label{eq:constraint} \\
q(E_R') & \geq  0~, \label{eq:geq} \\
\int dE_R' \, \frac{d\tilde{g}}{d E'_R} q(E_R') & = 0~. \label{eq:comp}
\end{align}
\Eq{eq:var} is the familiar equation resulting from varying the modified functional  with respect to $\tilde{g}$, and \Eq{eq:constraint} is the desired monotonicity constraint. \Eq{eq:comp} is a complementarity condition which ensures that the shift in $L$ vanishes on the solution, just as the extra Lagrange multiplier term vanishes on the solution in the case of equality constraints. When combined with \Eqs{eq:constraint}{eq:geq}, \Eq{eq:comp} enforces that at every point $E_R'$, either $d\tilde{g}/dE_R' = 0$ (saturating the inequality constraint), or $q(E_R') = 0$.

Suppose that the solution $\tilde{g}(E_R')$ has nonzero derivative at some point $E_0$. Then by \Eq{eq:comp}, $q(E_0)= 0$. Moreover, we must have $dq/dE_R'= 0$ at $E_0$ since if not, we would violate the positivity condition (\ref{eq:geq}) at $E_0 + \epsilon$ for arbitrary $\epsilon > 0$. Thus, \Eq{eq:var} becomes $\delta L / \delta \tilde{g} = 0$, or taking the functional derivative explicitly,\footnote{We have divided out by $K(E_R')$ which is legitimate so long as we are considering $(a,b) \in [E_{min}, E_{max}]$.}
\be
\sum_{i=1}^{N_O} \frac{G(E_i, E_0)}{\gamma_i} = 1,
\label{eq:contradiction}
\ee
where
\be
\gamma_i =  \mu_i + \int dE_R'' \, G(E_i, E_R'')K(E_R'')\tilde{g}(E_R'')~,
\label{eq:gammai}
\ee
is the total differential event rate at $E_i$. By property (ii), as a function of $E_0$, the left-hand side of (\ref{eq:contradiction}) is a sum of $N_O$ peaked functions, weighted by various factors $\gamma_i$. By property (iii), $G$ has no flat regions unless it vanishes, so a sum of functions of this form will cross a horizontal line at most $2N_O$ times; thus, only isolated points $E_0$ are solutions. This proves that $\tilde{g}(E_R')$ must be flat except at isolated points $E_0 = \widetilde{E}_j$; in other words, it is a sum of step functions.

To determine the number and position of the points $\widetilde{E}_j$, we can read \Eq{eq:var} as a differential equation for $q$:
\be
\frac{dq}{dE_R'} = K(E_R') \left (1 - \sum_{i=1}^{N_O} \frac{G(E_i, E_R')}{\gamma_i} \right ).
\label{eq:diffq}
\ee
The solution to this equation depends on the $\gamma_i$, which in turn depend on the full solution function $\tilde{g}(E_R')$, so we cannot integrate this equation directly. In fact this turns out not to be necessary. 

By the complementarity condition, we must have $q(\widetilde{E}_j) = 0$, but to preserve the positivity condition (\ref{eq:geq}), we must also have
\be
\left . \frac{dq}{dE_R'} \right |_{\widetilde{E}_j} = 0, \hspace{3mm} \left . \frac{d^2q}{dE_R'^2} \right |_{\widetilde{E}_j} \geq 0
\label{eq:qconditions}
\ee
at the roots $\widetilde{E}_j$ of $q$. Taking the derivative of \Eq{eq:diffq}, and using the assumption $dK/dE_R' \approx 0$, the condition on the second derivative becomes
\be
-\sum_{i=1}^{N_O} \frac{1}{\gamma_i} \left . \frac{\partial G(E_i, E_R')}{\partial E_R'} \right|_{E'_R = \widetilde{E}_j} \gtrsim 0.
\label{eq:convexity}
\ee
By property (ii), $G(E_i, E_R' )$ will be peaked at $E_R' = E_i$, so for $E_R'$ close but not equal to $E_i$, the $i$th term in the sum should dominate. The derivative of a peaked function is negative to the right of the peak, so to satisfy the inequality, we must have $\widetilde{E}_j > E_i$. In other words, the positions of the steps shift to the right slightly. 

The positions $\widetilde{E}_j$ are given by solving $\left.dq/dE_R' \right |_{\widetilde{E}_j}= 0$, which we already used as \Eq{eq:contradiction} to derive the shape of $\widetilde{g}$.  There are at most $2N_O$ solutions, but only $N_O$ of these solutions will satisfy the convexity condition (\ref{eq:convexity}) and could qualify as roots of $q$. For sufficiently large $\gamma_i$, the peak of $G(E_i, E_R')$ may dip below the line of height 1, so there may be fewer than $N_O$ solutions; the same is true if two peaks are close enough to one another to merge together. Furthermore, it may be the case that both conditions (\ref{eq:qconditions}) are satisfied at $\widetilde{E}_j$, but $q(\widetilde{E}_j) \neq 0$, in which case there would be no step at $\widetilde{E}_j$. We conclude that in the case of physically reasonable $G(E_R, E_R')$, \emph{the optimal halo integral $\tilde{g}(E_R)$ is given by a sum of at most $N_O$ step functions}. 

Rather than integrate \Eq{eq:diffq}, one may simply use this knowledge to perform at most a $2N_O$-parameter numerical minimization of the log-likelihood subject to the monotonicity constraint on $\tilde{g}$.

\bibliographystyle{JHEP}
\bibliography{astind}

\end{document}